\documentclass[sigconf]{acmart}

\usepackage{algorithm}
\usepackage{algorithmic}
\usepackage{amsmath}
\usepackage{multirow}
\usepackage{multicol}
\usepackage{balance}

\AtBeginDocument{%
  }

\copyrightyear{2025}
\acmYear{2025}
\setcopyright{acmlicensed}\acmConference[WWW Companion '25]{Companion Proceedings of the ACM Web Conference 2025}{April 28-May 2, 2025}{Sydney, NSW, Australia}
\acmBooktitle{Companion Proceedings of the ACM Web Conference 2025 (WWW Companion '25), April 28-May 2, 2025, Sydney, NSW, Australia}
\acmDOI{10.1145/3701716.3715215}
\acmISBN{979-8-4007-1331-6/2025/04}

\begin{document}

\title{Creator-Side Recommender System: Challenges, Designs, and Applications}

\author{Xiaoshuang Chen}
\authornote{These authors contributed equally to this research.}

\affiliation{%
  \institution{Kuaishou Technology}
  \city{Beijing}
  \country{China}
}
\email{chenxiaoshuang@kuaishou.com}

\author{Yibo Wang}
\authornotemark[1]

\affiliation{%
  \institution{Kuaishou Technology}
  \city{Beijing}
  \country{China}
}
\email{wangyibo10@kuaishou.com}

\author{Yao Wang}
\authornotemark[1]
\affiliation{%
  \institution{Kuaishou Technology}
  \city{Beijing}
  \country{China}
}
\email{wangyiyan@kuaishou.com}

\author{Husheng Liu}

\affiliation{%
  \institution{Kuaishou Technology}
  \city{Beijing}
  \country{China}
}
\email{liuhusheng@kuaishou.com}

\author{Kaiqiao Zhan}
\affiliation{%
  \institution{Kuaishou Technology}
  \city{Beijing}
  \country{China}
}
\email{zhankaiqiao@kuaishou.com}

\author{Ben Wang}\authornote{Corresponding author.}
\affiliation{%
  \institution{Kuaishou Technology}
  \city{Beijing}
  \country{China}
}
\email{wangben@kuaishou.com}

\author{Kun Gai}
\affiliation{%
  \institution{Unaffiliated}
  \city{Beijing}
  \country{China}
}
\email{gai.kun@qq.com}
\renewcommand{\shortauthors}{Xiaoshuang Chen et al.}

\begin{abstract}
Users and creators are two crucial components of recommender systems. Typical recommender systems focus on the user side, providing the most suitable items based on each user’s request. In such scenarios, a few items receive a majority of exposures, while many items receive very few. This imbalance leads to poorer experiences and decreased activity among the creators receiving less feedback, harming the recommender system in the long term. To this end, we develop a creator-side recommender system, called \textbf{DualRec}, to answer the following question: \textbf{How to find the most suitable users for each item to enhance the creators' experience?} We show that typical user-side recommendation algorithms, such as retrieval and ranking algorithms, can be adapted into the creator-side versions with just a few modifications. This greatly simplifies algorithm design in DualRec. Moreover, we discuss a unique challenge in DualRec: the user availability issue, which is not present in user-side recommender systems. To tackle this issue, we incorporate a user availability calculation (UAC) module to effectively enhance DualRec's performance. DualRec has already been implemented in Kwai, a short video recommendation system with over 100 million users and over 10 million creators, significantly improving the experience for creators.
\end{abstract}

\ccsdesc[500]{Information systems~Recommender systems}

\keywords{Recommender System, Creator Experience, Cold Start}


\maketitle
\vspace{-2mm}
\section{Introduction} \label{sec:introduction}

\begin{table*}
\caption{Comparisons between user-side and creator-side recommender systems.}
\vspace{-2mm}
\centering
\begin{tabular}{c|c|c}
    \hline\hline
     & \textbf{User-Side Recommender System} & \textbf{Creator-Side Recommender System (DualRec)}  \\
    \hline
    Request Source & User $u\in\mathcal{U}$ & Item $i\in\mathcal{C}$ \\
    \hline
    Candidate Set & the item set $\mathcal{C}$ & the user set $\mathcal{U}$ \\
    \hline
    Results & an item set $\mathcal{C}_u$ to be recommended to User $u$ & a user set $\mathcal{U}_i$ who will receive Item $i$ \\
    \hline
    Objective & user experience on the recommended items & creator experience when receiving feedback from the users \\
    \hline\hline
\end{tabular}
\label{table:comparisons-user-item}
\end{table*}

\begin{figure}[t]
    \centering
    \includegraphics[width=\columnwidth, trim=0 5 0 0, clip]{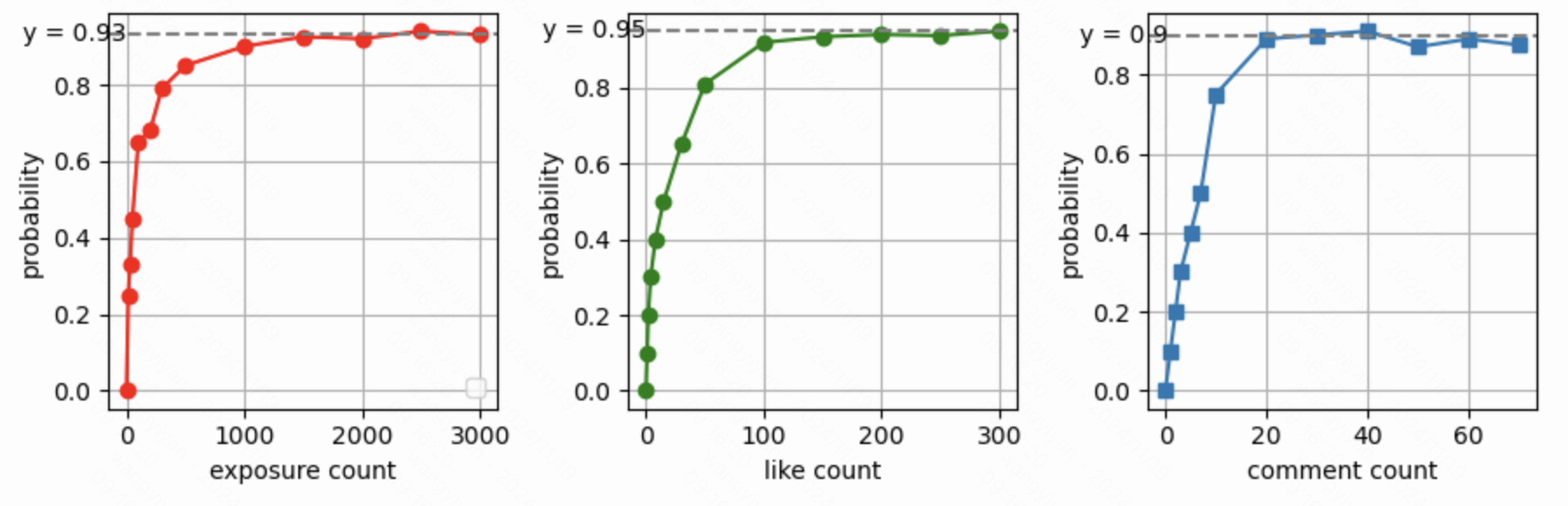}
    \caption{Impacts of Engagement on Future Post Probability.}
    \label{fig:intro-impact}
\vspace{-5mm}
\end{figure}

\begin{figure*}[t]
    \centering
    \includegraphics[width=0.75\linewidth,trim=0 10 0 0, clip]{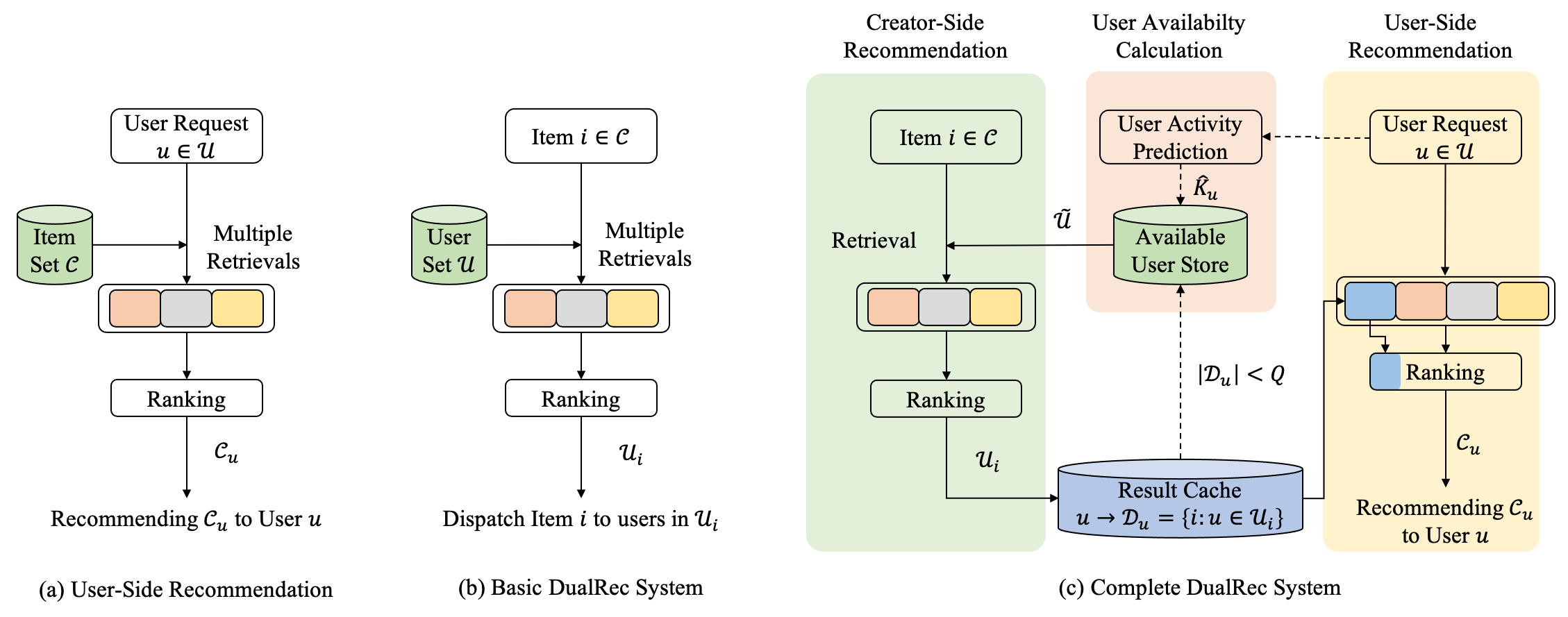}
    \caption{The DualRec Framework.}
    \label{fig:dualrec}
\vspace{-5mm}
\end{figure*}

Recommender systems are widely employed in areas such as short video platforms \cite{tang2017popularity,wu2018beyond}, search engines \cite{huang2020embedding,searchSi,searchShi}, and E-commerce \cite{gu2020hierarchical,linden2003amazon,zhou2018deep,aliXiao} to connect users with relevant content. Typically, these systems aim to provide the most relevant items for each user request to enhance user satisfaction. We refer to these systems as \textbf{user-side recommender systems}. User-side recommendations often result in a limited number of items receiving significant exposure, while many creators and their content remain unseen by consumers.
Receiving fewer exposures and less feedback typically results in poorer experiences and decreased activity among creators.

Recently, there has been a growing body of literature on dispatching long-tail items, such as Google's cold start system \cite{WangKDD23Fresh}, FairRec \cite{patro2020fairrec}, and others \cite{zhang2021model,zhu2021fairness,zhang2023empowering,yuan2023go}. These approaches create more opportunities for underexposed items, but they still prioritize user satisfaction when distributing these underexposed items. 
In a real-world recommender system, users and content creators respond differently to various types of feedback. For instance, in Kwai, a short video platform, a user's watch time on a video is crucial for reflecting user satisfaction; however, this watch time is not visible to the video's creator and, therefore, cannot impact the creator's satisfaction. In contrast, creators tend to feel more satisfied when they receive likes and comments on their videos. As shown in Figure \ref{fig:intro-impact}, interactions such as likes and comments received by creators have a significant positive incentive effect on their content creation. This disparity between user and creator satisfaction presents challenges for recommender systems to consider the creator experience.

To this end, this paper discusses the development of a creator-side recommender system aimed at \textbf{finding the most appropriate users for each item to enhance creator satisfaction}. This creator-side recommender system is called \textbf{DualRec}, where ``dual'' signifies its correspondence with user-side recommender systems. DualRec is a mirror system to user-side recommender systems, with all concepts from the user-side recommender system having corresponding concepts in DualRec. Table \ref{table:comparisons-user-item} presents a comparison between user-side recommender systems and DualRec.

We first demonstrate that there are many similarities between traditional user-side recommender systems and DualRec. Specifically, a typical user-side recommender system, illustrated in Figure \ref{fig:dualrec}(a), consists of multiple retrieval modules that generate candidate item sets, along with a ranking module that predicts user feedback on the items for final recommendations. Similarly, DualRec, as shown in Figure \ref{fig:dualrec}(b), also includes retrieval and ranking modules. Furthermore, we illustrate that typical algorithms used in traditional user-side recommender systems can be adapted into their corresponding creator-side versions with just a few modifications. This feature ensures the compatibility of DualRec with user-side recommendation systems, allowing research related to the latter to be directly applied to the creator-side recommendation system.

We then demonstrate that, unlike user-side recommender systems, creator-side recommender systems face the \textbf{user availability issue}. In user-side recommender systems, when we want to recommend an item for a user request, we can generally find the item as long as it hasn't been deleted by the author. However, in creator-side recommendation systems, when we seek to identify suitable users for an item, there is no guarantee that the users will receive it. On one hand, users may simply not visit the platform. For instance, the number of daily active users is about 30\% of the total user base on the Kwai platform. On the other hand, there are no constraints on the number of items assigned to the same user, and this allows for many items to be allocated to one user, which contrasts with the reality that a single user cannot consume too many items. To address this issue, we introduce a user availability calculation(UAC) module to calculate and maintain available users and use UAC to guide the creator-side recommendations, as shown in Figure \ref{fig:dualrec}(c).

In summary, our contributions are as follows: 
\begin{itemize}
    \item We introduce a novel creator-side recommender system called DualRec, designed to find the most appropriate users for each item to enhance creator satisfaction, which cannot be covered by traditional user-side recommender systems. 
    \item We discuss the detailed implementation of DualRec, and show that existing user-side recommendation algorithms can be mirrored to the creator-side version with a few changes, which simplifies the implementation of DualRec.
    \item We address the user availability issue, a unique challenge in creator-side recommendation systems, and introduce UAC to overcome this challenge.
    \item We implement DualRec on the Kwai platform, showing that DualRec significantly enhances the creator experience.
\end{itemize}

\section{Problem Formulation} \label{sec:problem-formulation}
Before introducing the creator-side recommendation, we first formally discuss the user-side recommendation.  Denote the user set as $\mathcal{U}$ and the item set as $\mathcal{C}$. When a user $u\in\mathcal{U}$ sends a request to the system, the user-side recommender system returns $K$ items to the user from the item set $\mathcal{C}$, aiming to enhance user satisfaction. We denote the user-side recommendation process as follows:
\begin{equation} \label{eq:user-side-recommendation}
    \mathcal{C}_u = \mathcal{R}^U\left(u, \mathcal{C},K\right)
\end{equation}
where $\mathcal{C}_u$ is the returned item set with a size of $K$, and $\mathcal{R}^U$ represents an abstract model of the user-side recommendation process. Typically, the recommender system must provide real-time recommendations from a large item set $\mathcal{C}$ for each user request. Hence, the implementation of $\mathcal{R}^U$ involves a complex infrastructure that includes retrieval and ranking modules, as shown in Figure \ref{fig:dualrec}(a).

The creator-side recommendation is a mirror problem of the user-side recommendation. Given an item $i\in\mathcal{C}$ uploaded by a certain creator, the creator-side recommender system dispatches it to $L$ users from the user set $\mathcal{C}$, aiming to enhance creator satisfaction. Like Eq. \eqref{eq:user-side-recommendation}, we write the creator-side recommendation process as:
\begin{equation} \label{eq:item-side-recommendation}
    \mathcal{U}_i = \mathcal{R}^C\left(i, \mathcal{U},L\right)
\end{equation}
where $\mathcal{U}_i$ is the target user set with size $L$ for Item $i$, and $\mathcal{R}^C$ is an abstract representation of the creator-side recommendation.
We aim to discuss the specific structure of $\mathcal{R}^C$, and
there exist two kinds of problems:
\begin{itemize}
    \item What components are needed in the creator-side recommender system? What are the similarities and differences of these components between the user-side recommendation system and the creator-side recommendation system?
    \item How to address the \textbf{user availability issue}? In other words, how can we ensure that the system returns users who can actually consume the item, meaning these users will visit the platform and will not be overwhelmed by too many items?
\end{itemize}

To tackle these issues, we propose a novel DualRec system. Specifically, we introduce a basic version of DualRec in Section \ref{sec:mirror} to tackle the first issue. Then, we discuss the solution to the user availability issue and introduce the complete DualRec system in Section \ref{sec:dualrec}. 

\section{DualRec: Mirror of User-Side Recommenders} \label{sec:mirror}
This section presents the basic DualRec system in Figure \ref{fig:dualrec}(b), which is a mirror of the user-side recommendation. We discuss the implementation of the retrieval and ranking processes, showing that user-side recommendation algorithms can be easily adapted into their corresponding creator-side versions with a few modifications.

\subsection{System Overview}
We first discuss a typical user-side recommender system consisting of retrieval and ranking modules, as shown in Figure \ref{fig:dualrec}(a). Specifically, the retrieval module employs multiple lightweight algorithms to extract a small set of candidate items from the item set $\mathcal{C}$, while the ranking module makes more precise predictions regarding user feedback on these items to select the final items.

The DualRec framework, illustrated in Figure \ref{fig:dualrec}(b), mirrors the user-side recommender system. It includes creator-side retrieval and ranking modules. When a new item $i$ is uploaded by a creator, the creator-side retrieval module finds candidate users from the user set $\mathcal{U}$, and the creator-side ranking module then selects the final user set $\mathcal{U}_i$ from the candidate user set from the retrieval module based on more accurate predictions to enhance creator satisfaction.

Our fundamental idea of the algorithm design in DualRec is to treat users and items as two symmetrical components of the recommendation system. By interchanging users and items in the user-side recommendation algorithms, we can derive the corresponding creator-side versions. We are now prepared to present a detailed implementation of the retrieval and ranking modules.

\subsection{Retrieval} \label{sec:dualrec:retrieval}
 
Typical user-side retrievals include similarity-based retrievals \cite{he2017neural,wang2018billion,chang2021sequential}, and model-based retrievals\cite{huang2013learning,ModelEk,electronics11010141}. Here we discuss the corresponding creator-side versions of these retrievals.

\subsubsection{Similarity-Based Retrievals}
\begin{figure}[t]
    \centering
    \includegraphics[width=\columnwidth,trim=0 10 0 0, clip]{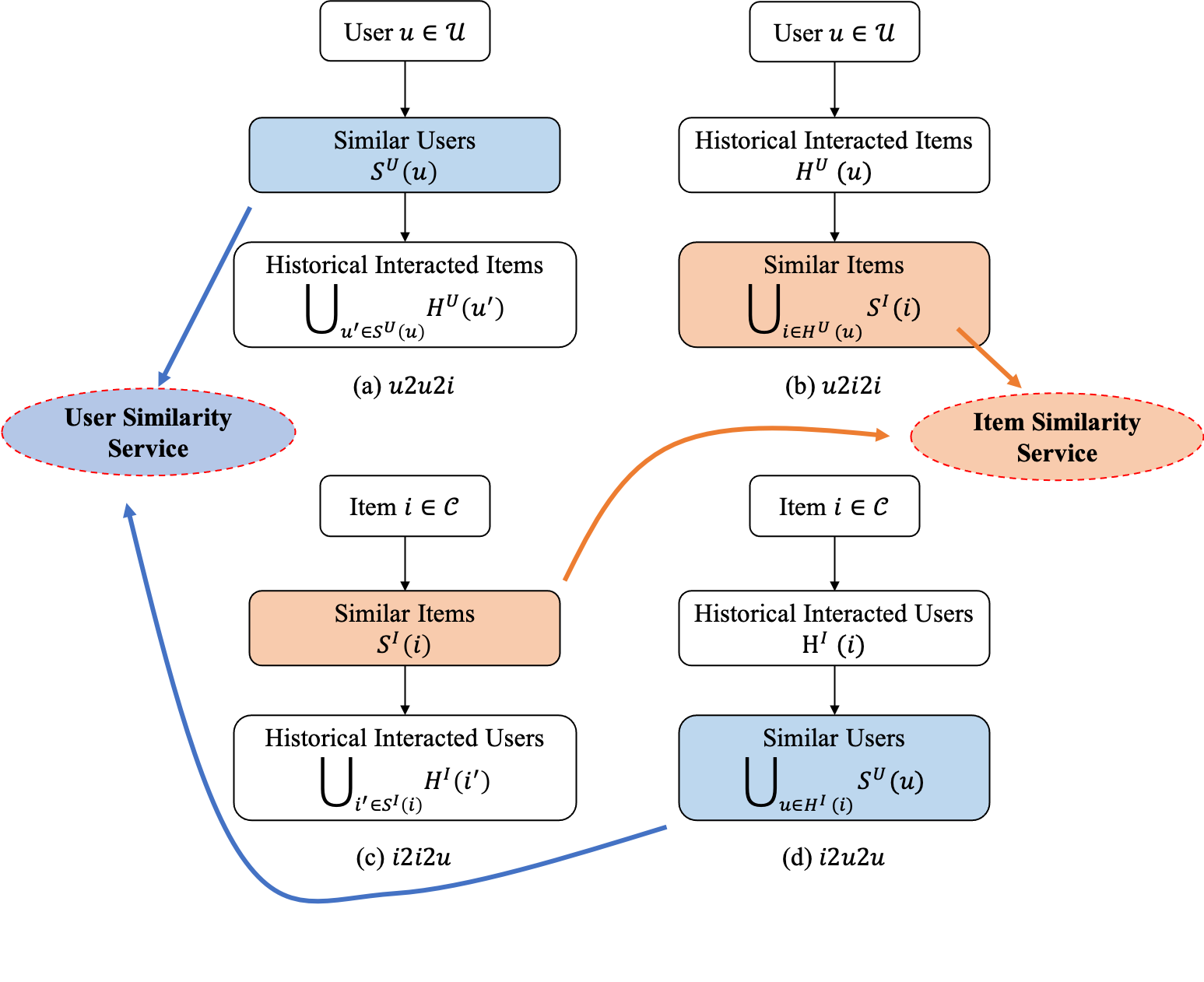}
    \caption{Similarity-based retrievals: (a) $u2u2i$; (b) $u2i2i$; (c) $i2i2u$; (d) $i2u2u$.}
    \label{fig:similarity-based-retrieval}
\vspace{-5mm}
\end{figure}

Similarity-based retrievals attempt to provide recommendations to users/items based on similar users or items. We first discuss the similarity-based retrieval algorithms in user-side recommender systems. Specifically, a similarity calculation algorithm calculates the similarity among users or items. Typical similarity calculation algorithms include the collaborative-filtering-based algorithms \cite{CFSarwar,he2017neural} and graph-based algorithms \cite{wang2018billion,chang2021sequential,graphHu,graphDe}. Then, similarity services are built to find similar users/items of a certain user/item. There are two kinds of similarity services, i.e., the user similarity service, denoted by $S^U$, and the item similarity service, denoted by $S^I$. Based on the similarity service, there are also two kinds of similarity-based retrievals in the user-side recommender systems, namely the user-side user-similarity-based retrieval (Figure \ref{fig:similarity-based-retrieval}(a), denoted by $u2u2i$) and the user-side item-similarity-based retrieval (Figure \ref{fig:similarity-based-retrieval}(b), denoted by $u2i2i$). When a user request $u$ comes, the $u2u2i$ calls the user similarity service to find similar users $S^U(u)$ and then retrieves the previous interacted items of the users in $S^U(u)$. Formally we have:
\begin{equation} \label{eq:u2u2i}
    \mathcal{C}_u^{u2u2i} = \left\{H^U(u')|u'\in S^U(u)\right\}
\end{equation}
where $H^U(u')$ is the previous positively interacted items of the user $u'$. In contrast, the $u2i2i$ calls the item similarity service to find similar items of User $u$'s previous interacted items:
\begin{equation} \label{eq:u2i2i}
    \mathcal{C}_u^{u2i2i} = \bigcup \left\{S^I(i')|i'\in H^U(u)\right\}
\end{equation}

By simply exchanging all the concepts of users and items in the user-side similarity-based retrievals, we obtain the corresponding creator-side similarity-based retrievals. Specifically, the creator-side similarity-based retrievals can be classified into the creator-side item-similarity-based retrievals $i2i2u$, which is the mirrored version of $u2u2i$, and the creator-side user-similarity-based retrievals $i2u2u$, which is the mirrored version of $u2i2i$. We write the two kinds of retrievals as
\begin{equation} \label{eq:i2i2u-i2u2u}
    \begin{aligned}
        \mathcal{U}_i^{i2i2u} &= \left\{H^I(i')|i'\in S^I(i)\right\} \\
        \mathcal{U}_i^{i2u2u} &= \bigcup \left\{S^U(u')|u'\in H^I(i)\right\}
    \end{aligned}
\end{equation}
where $H^I(i)$ is the previous positively interacted users of the item $i$. The $i2i2u$ and $i2u2u$ algorithms are shown in Figure \ref{fig:similarity-based-retrieval}(c) and (d).

The similarity calculation algorithms and the similarity services are crucial to the performance of similarity-based retrievals, and a key finding here is that the similarity calculation algorithms in user-side similarity-based retrievals can be directly applied to the creator-side similarity-based retrievals. Specifically, by comparing the creator-side similarity-based retrieval in Eq. \eqref{eq:i2i2u-i2u2u} with the user-side similarity-based retrieval in Eq. \eqref{eq:u2u2i}\eqref{eq:u2i2i}, we find that the user similarity service $S^U$ in $i2u2u$ is the same as the similarity service used in $u2u2i$, and the similarity service $S^I$ in the $i2i2u$ is the same as that used in $u2i2i$. Therefore, there is no need to train extra models for creator-side similarity-based retrievals.

\subsubsection{Model-Based Retrievals}
\begin{figure}[t]
    \centering
    \includegraphics[width=0.9\columnwidth,trim=0 10 0 0, clip]{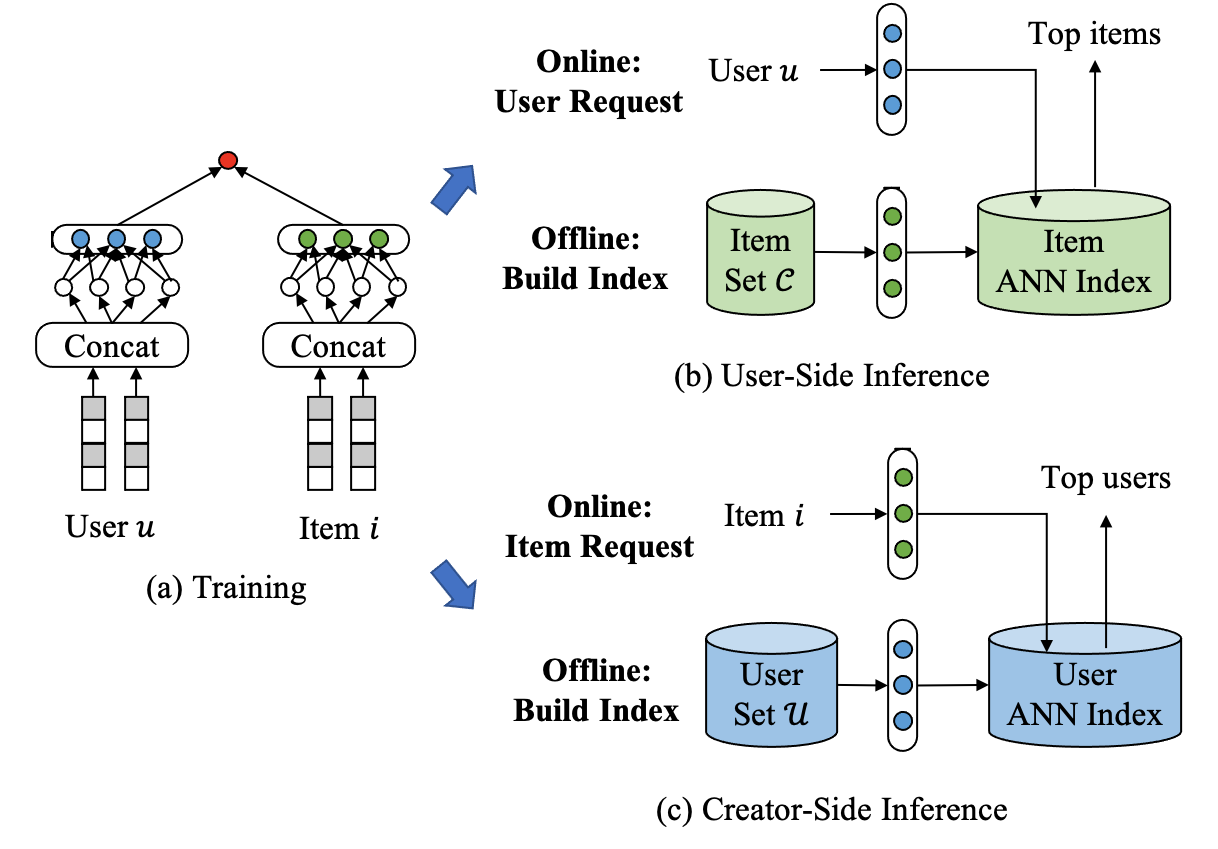}
    \caption{Two-tower model retrieval}
    \label{fig:two-tower-retrieval}
\vspace{-5mm}
\end{figure}

Here we discuss the most popular model-based retrieval, i.e., the two-tower model retrieval. The two-tower model retrieval contains a user tower, which outputs the user embedding $\boldsymbol{e}^u$, and an item tower, which outputs the item embedding $\boldsymbol{e}_i$, and regards the inner product of the user and item embeddings as the logit. Formally we have
\begin{equation} \label{eq:two-tower-training}
    \begin{aligned}
        \boldsymbol{e}_u = f^U\left(\boldsymbol{x}_u\right),&\boldsymbol{e}_i = f^I\left(\boldsymbol{x}_i\right) \\
        logit =& \boldsymbol{e}_u\cdot \boldsymbol{e}_i
    \end{aligned}
\end{equation}
where $f^U$ and $f^I$ are the user tower and the item tower, respectively. $\boldsymbol{x}_u$ and $\boldsymbol{x}_i$ are the user features and the item features, respectively. In the training process, the $logits$ are supervised by the users' feedback on the items, as shown in Figure \ref{fig:two-tower-retrieval}(a). In the inference process of the user-side two-tower model retrieval, an offline index builder traverses all items $i\in\mathcal{C}$, computing the item embedding $\boldsymbol{e}_i$, and then builds an approximated nearest neighbor (ANN) service \cite{wieschollek2016efficient}. When a user $u$ sends a request, the user embedding $\boldsymbol{e}_u$ is calculated online, and the $\boldsymbol{e}_u$ is used to search the most relevant items from the ANN server, as shown in Figure \ref{fig:two-tower-retrieval}(b):
\begin{equation}
    \mathcal{C}_u^{two-tower} = \textbf{Top}_{i\in\mathcal{C}}\left\{\boldsymbol{e}_u\cdot\boldsymbol{e}_i\right\}
\end{equation}
The ANN service makes it possible to retrieve $\mathcal{C}_u^{two-tower}$ from a very large candidate set $\mathcal{C}$ in a low latency, which is necessary for the two-tower model retrieval.

The creator-side two-tower retrieval is also the mirrored version of the user-side two-tower retrieval. According to Eq. \eqref{eq:two-tower-training}, the users and the items are symmetric in the two-tower model, which means the training process of the creator-side two-tower model can be the same as the user-side two-tower model in Figure \ref{fig:two-tower-retrieval}(a). However, the inference processes of user-side and creator-side two-tower retrievals are different. We must build an ANN index according to the embeddings $\boldsymbol{e}_u$ of all the users. Then, when an item request comes, the retrieval algorithm returns the most relevant users from the ANN server, as shown in Figure \ref{fig:two-tower-retrieval}(c):
\begin{equation}
    \mathcal{U}_i^{two-tower} = \textbf{Top}_{u\in\mathcal{U}}\left\{\boldsymbol{e}_u\cdot\boldsymbol{e}_i\right\}
\end{equation}

Therefore, by replacing the item ANN index with the user ANN index, we derive the creator-side two-tower retrieval algorithm.

\subsection{Ranking}
The ranking module receives the candidate set from the retrieval module and then determines the final output of the recommendation process. The ranking module contains two components, i.e., the prediction module and the decision module. The prediction module uses deep models to predict the user's multiple feedback, e.g., the watch time, likes, shares, etc. on the items. Then, the decision module uses these predictions to determine the final outputs. Formally, assume there are $M$ types of feedback to be considered, and denote the $m$-th prediction score of User $u$ and Item $i$ as $s_{ui}^m$. Then, the prediction module generates the predictions $s_{ui}^m$, and the decision module calculates the integrated score as:
\begin{equation} \label{eq:user-side-score-integration}
    f_{ui} = \sum_{m=1}^M\alpha_ms_{ui}^m
\end{equation}
where $\alpha_m$ is the coefficient of the prediction of the $m$-th feedback. Then the user-side ranking module returns the top-$K$ items according to the integrated score $f_{ui}$.

The creator-side ranking module has the same structure as the user-side ranking module except for two detailed settings. First, the model structures of the prediction models are different because DualRec focuses more on the model performance on each item, especially the long-tail items lacking training samples. Second, the ranking coefficient of each prediction usually differs between the creator-side and user-side recommender systems.

\subsubsection{Prediction Models} \label{sec:mirror:ranking:prediction}
\begin{figure}[t]
    \centering
\includegraphics[width=0.75\columnwidth,trim=0 0 0 0, clip]{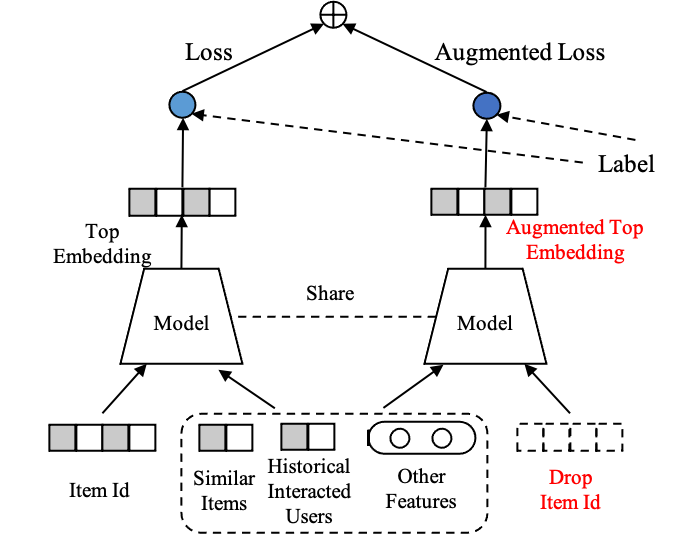}
    \caption{The structure of the prediction model.}
    \label{fig:model-structure}
\vspace{-5mm}
\end{figure}

Typical prediction models are subject to popularity bias. They fail to provide accurate predictions on long-tail items which making up a significant proportion of the creator-side recommender system. To address this, we make the following changes to the prediction model, as shown in Figure \ref{fig:model-structure}:
\begin{enumerate}
    \item We augment the training sample by dropping the item ID feature, and train the model with both the original sample and the augmented sample to force the model to extract information from side features other than the item IDs.
    \item We add more item-side features, such as the users who have recently interacted with the item and the information on similar items according to video understanding models.
\end{enumerate}
We will discuss the impact of these changes in the experiments.

\subsubsection{Score Integration} \label{sec:mirror:ranking:score-integration}
We use the following formula:
\begin{equation}\label{eq:item-side-score-integration}
    g_{ui}=\sum_{m=1}^M\beta_ms_{ui}^m / \theta_u
\end{equation}
There are two differences between
Eq. \eqref{eq:user-side-score-integration} and Eq. \eqref{eq:item-side-recommendation}. First, there is a bias term $\theta_u$, which means different users have different values in DualRec. This is because we find that different users' feedback usually leads to different future exposures, and then leads to different creator satisfaction. Second, the coefficients of the users' explicit interactions on items, e.g., likes and comments, are more important in DualRec than in user-side recommendations because the items' creators can receive these interactions and interact with the users. In contrast, the coefficients of the watch time are less important because the creators cannot sense the watch time of users.

\section{Tackling the User Availability Issue} \label{sec:dualrec}
We have discussed the retrieval and ranking algorithms of DualRec. However, a key difference between the creator-side recommendation and the user-side recommendation is the user availability issue as discussed in Section \ref{sec:problem-formulation}. We discuss the solution to this issue, and provide the complete DualRec framework, as shown in Figure \ref{fig:dualrec}(c).

\subsection{The Complete DualRec Framework}
The key motivation is to maintain an available user set $\widetilde{\mathcal{U}}$ as a subset of the total user set $\mathcal{U}$. We define the available users as those who will visit the platform in an acceptable period (one day in our platform) and have not been dispatched with too many items. Then, the DualRec system finds target users in the available user set for each item and then updates the available user set to prepare for the next item request. With the available user set, the recommendation results of DualRec can get rid of the user availability problem.

Formally, we define some notations:
\begin{itemize}
    \item The user activity state $a_u \in \{0,1\}$: $a_u=1$ means the user will visit the platform within an acceptable period, and vice versa.
    \item The user matching set $\mathcal{D}_u$: the set of items which have been matched to the user $u$. Formally we have:
    \begin{equation} \label{eq:d-u}
        \mathcal{D}_u = \left\{i:u\in \mathcal{U}_i\right\}
    \end{equation}
\end{itemize}
Then, the available user set $\widetilde{\mathcal{U}}$ can be defined as:
\begin{equation} \label{eq:available-user-set}
    \widetilde{\mathcal{U}} = \left\{u\in\mathcal{U}:a_u = 1,\left|\mathcal{D}_u\right|<Q\right\}
\end{equation}
where $Q$ is the maximum number of items the user can consume. Although different users can consume different numbers of items, we set $Q$ as a constant number for simplicity.

Based on these notations, we provide the complete DualRec framework, as depicted in Figure \ref{fig:dualrec}(c). The complete DualRec contains a UAC module, a basic DualRec module discussed in Section \ref{sec:mirror}, a result cache, and a combined recommendation module. The UAC module consists of a user activity prediction module, which predicts the user's future activity $a_u$, and an available user store, which maintains the available user set $\widetilde{\mathcal{U}}$ according to Eq. \eqref{eq:available-user-set}. Based on the available user set, the basic DualRec continuously recommends users to each item, and the results $\mathcal{U}_i$ is used to update the result cache and the available user store. Users with enough matched items in $\mathcal{D}_u$ will be eliminated from the available user set $\widetilde{\mathcal{U}}$. Finally, the combined recommendation module combines the items in the match set $\mathcal{D}_u$ and the user-side recommendation results when User $u$ visits the platform. The detailed algorithm is provided in Algorithm \ref{alg:dualrec}.
\begin{algorithm}
\caption{The Algorithm of Complete DualRec}
\label{alg:dualrec}
\begin{algorithmic}[1]
\STATE Input: the user set $\mathcal{U}$, the item set $\mathcal{C}$.

\STATE Output: the target user set $\mathcal{U}_i$ for each item $i\in\mathcal{C}$.
\STATE
\STATE \textbf{Initialize}
\FOR{each user $u$ in $\mathcal{U}$}
    \STATE The user activity prediction module predicts the user's future activity $a_u$.
\ENDFOR
\STATE The UAC the available user set $\widetilde{\mathcal{U}}$ as $\widetilde{\mathcal{U}} = \left\{u\in\mathcal{U}:a_u = 1\right\}$.

\STATE
\STATE \textbf{Creator-side recommendation}
\FOR{each item $i\in\mathcal{C}$}
        \STATE A basic DualRec algorithm runs the recommendation process in the available user set $\widetilde{\mathcal{U}}$: $\mathcal{U}_i = \mathcal{R}^I\left(i,\widetilde{U}, L\right)$
        \STATE  Update the user matching set for $u\in\mathcal{U}_i$:
        $\mathcal{D}_u \gets \mathcal{D}_u \cup \{i\}$,
    and then store the updated $\mathcal{D}_u$ in the result cache.
    \STATE We update the available user set $\widetilde{U}$ according to the updated $\mathcal{D}_u$ and Eq. \eqref{eq:available-user-set}.
\ENDFOR

\end{algorithmic}
\end{algorithm}

The proposed framework can effectively solve the user availability issue. Firstly, the user activity prediction module chooses the users who are more likely to visit the platform. Secondly, the rule of the available user set $\widetilde{\mathcal{U}}$ in Eq. \eqref{eq:available-user-set} eliminates the users who have been dispatched with enough items.
In the next subsection, we will discuss the detailed implementation of these modules.

\subsection{Detailed Implementation}
\subsubsection{User Activity Prediction} \label{sec:dualrec:user-activity-prediction}
We train an XGBoost model\cite{chen2016xgboost} to estimate each user's daily activity according to the user's previous activity. The input features include the user's previous activity in the last 30 days, the day of the week, the user's accumulated interaction with the platform, and the output score is the probability that the user will visit the platform on the next day. We run the user activity prediction module daily, which traverses all the users $u\in\mathcal{U}$, and predicts the users' activity $a_u$ in the next day.

\subsubsection{Updating the available user set $\widetilde{U}$} We use an available user store to maintain the available user set $\widetilde{\mathcal{U}}$. When an item $i$ is uploaded by its creator, the DualRec system returns the user set $\mathcal{U}_i$ for the item $i$. Then, we update the user matching sets of the users in $\mathcal{U}_i$ by adding the item $i$, i.e.,
\begin{equation}
    \mathcal{D}_u\gets \mathcal{D}_u \cup \{i\}
\end{equation}
Then, if $\left|\mathcal{D}_u\right|\geq Q$, which means the user $u$ has been matched with enough items, we eliminate User $u$ from the available user set $\widetilde{\mathcal{U}}$.
\subsubsection{Combined Recommendation} \label{sec:dualrec:combination}
DualRec is crucial to improve the creators' experience, but it usually cannot replace the user-side recommendation because the users' experience is also very important. We use a combined recommendation module to integrate the creator-side and user-side recommendations. Specifically, the output of the creator-side recommender system is first put into a result cache, which stores the user matching set $\mathcal{D}_u$ defined in Eq. \eqref{eq:d-u}. Then, when User $u$ visits the platform, a user-side recommendation process will be performed, of which the cached user matching set $\mathcal{D}_u$ is one of the retrieval algorithms. Therefore, $\mathcal{D}_u$ and items from other retrievals are compared in the user-side ranking module. To ensure the exposures of items in $\mathcal{D}_u$, we add an extra value of these candidate items on their integrated scores $f_{ui}$ according to the item-side integrated score $g_{ui}$, i.e.
\begin{equation}
\label{eq:double-side-lamba}
    f'_{ui} = \left\{
    \begin{matrix}
        f_{ui} + \lambda g_{ui}, i\in\mathcal{D}_u \\
        f_{ui}, otherwise
    \end{matrix}
    \right.
\end{equation}
where $\lambda$ is the boosting coefficient. Such a setting increases the probability of the item-side recommendation results being returned to User $u$, which ensures the performance of DualRec.

\section{Live Experiments}
We implement DualRec in Kwai, a short video platform with over 100 million users and over 10 million creators. Creators upload over 4 million videos daily. We aim to answer the following questions:
\begin{itemize}
\item \textbf{Q1}: How to design an experimental framework to measure the impact of different strategies on creator satisfaction?
\item \textbf{Q2}: Does DualRec improve the creator satisfaction?
\item \textbf{Q3}: How does the proposed UAC module in Section \ref{sec:dualrec} solve the user availability issue?
\item \textbf{Q4}: What are the influences of different components in Sections \ref{sec:mirror} on the performance of DualRec?

\end{itemize}

\subsection{Experiment Design (Q1)}
\subsubsection{User-Creator Co-Diverted A/B Testing}
\begin{figure}[t]
    \centering
    \includegraphics[width=\columnwidth, trim=0 30 0 0, clip]{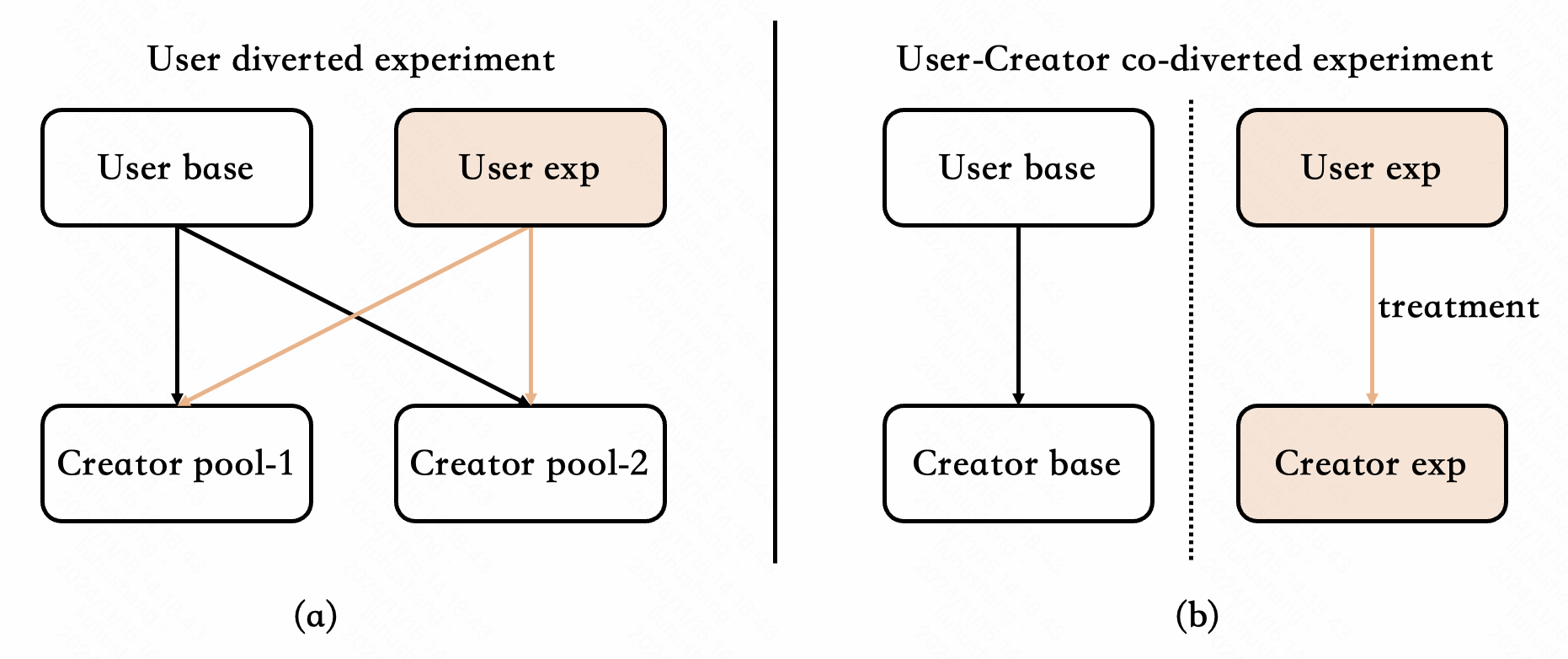}
    \caption{User diverted vs. User-Creator co-diverted}
    \label{fig:user-creator-exp}
\vspace{-5mm}
\end{figure}
In traditional A/B testing, we typically use a user-diverted setup, as shown in Figure \ref{fig:user-creator-exp}(a), where users are randomly assigned to either the control or experimental groups and receive corresponding recommendations from the entire creator pool. This allows for comparing user metrics, such as click-through rate (CTR) and dwell time, between the two groups, measuring the treatment effect from the user's perspective. However, since both groups share the same creators, the user-diverted setup cannot effectively assess any treatment effects on the creators due to treatment leakage. For example, an item exposed in the experiment may also appear in the control group. 

To address this issue, we adopt a user-creator co-diverted A/B testing setup (as shown in Figure \ref{fig:user-creator-exp}(b)), similar to the approach in \cite{WangKDD23Fresh}. In this setup, we first assign a proportion of creators to the control group and a separate, non-overlapping set of creators to the treatment group. Users are then proportionally assigned to the different groups, ensuring that users in the control group receive items only from the control creators, while users in the treatment group receive items only from the treatment creators. The user-creator co-diverted A/B Testing enables us to measure the treatment effect from the creator's perspective.

\subsubsection{Implementation Details}
The DualRec system is requested when a new item is uploaded by a certain creator and returns 200 users for each item. We allocate 2\% of the exposures for DualRec outputs while maintaining the others from the user-side recommendation to balance user and creator satisfaction.
We iteratively implement all the methods discussed in Sections \ref{sec:mirror} and \ref{sec:dualrec}:
\begin{itemize}
    \item \textbf{DualRec-v1}: A Complete DualRec system with necessary parts, i.e., a two-tower retrieval, a ranking module without the modifications discusses in \ref{sec:mirror:ranking:prediction}, a UAC module, and a combined recommendation module, as shown in Figure \ref{fig:dualrec}(c). In the two-tower-model-based retrieval, the positive samples are the users who have positive feedback, e.g. likes and comments, to the items, while the negative samples are randomly sampled from the user set $u\in\widetilde{\mathcal{U}}$. In the ranking module, we also adopt the two-tower model as a backbone. We predict the users' likes, comments, follows, shares, profile visits, and effective views on the items, and the integrated score is a weighted summation of these prediction scores. We set $Q=10$ for the UAC in Figure \ref{fig:dualrec}(c).
    \item \textbf{DualRec-v2}: Applying the modifications in Section \ref{sec:mirror:ranking:prediction} to the ranking models.
    \item \textbf{DualRec-v3}: Adding a content-based similarity $i2i2u$ retrieval to the DualRec system. Specifically, a content understanding algorithm outputs an embedding with dimension 128 to calculate the similarity between two items.
    \item \textbf{DualRec-v4}: Adding a $i2u2u$ retrieval based on a pre-trained user similarity model to the DualRec system.
    \item \textbf{DualRec-v5}: Modifying the score integration in the ranking module of DualRec with skills in Section \ref{sec:mirror:ranking:score-integration}. Introduce a user bias $\theta_u$, which is the user's average interaction rate.
\end{itemize}

\subsubsection{Baseline}
The baseline before launching \textbf{DualRec-v1} is a traditional user-side recommender system with extra strategies on new items with few exposures. The idea of these strategies is similar to FairRec\cite{patro2020fairrec}, which gives priority to items with fewer exposures. Specifically, there are extra retrievals, which only return new items with few exposures to each user request. Moreover, there is a boosting process in the ranking module, i.e., the scores of new items will be larger than the scores of old items when all predictions are the same. We denote this baseline as \textbf{Base}. Then we iteratively launched \textbf{DualRec-v1} to \textbf{DualRec-v5}, and each algorithm takes the previous version as the baseline.

\subsubsection{Metrics}
According to the setting of user-creator co-diverted A/B testing, there are two kinds of metrics, i.e., the user-side metrics and the creator-side metrics. The user-side metrics measure user satisfaction, including daily active users and users' daily accumulated watch time. It is out of this paper's scope to enhance user satisfaction, and these user-side metrics are regarded as constraints of the DualRec system. All of the experiments in this section do not decrease the user satisfaction measured by these metrics, and we will not list the user-side metrics for simplicity.

For the creator-side metrics, the core metric is the total \textbf{Daily Active Creator} (DAC), i.e. the number of creators who upload at least one item in one day. However, The increase in DAC is difficult to observe in a single experiment because it usually takes a considerable amount of time for the creators' activity levels to rise gradually. Therefore, we also consider a more sensitive metric positively related to the DAC, i.e. the number of the creators' new items reaching $E$ exposures, denoted by \textbf{ExpoReach-$E$}, where $E$ is set to different values, namely 1K/5K/10K, where the ``K'' means ``kilo''. Note that DualRec only dispatches 200 users to each item, which is much smaller than $E$; therefore, an increase of ExpoReach-$E$ means that the user-side recommender system also discovers these items better than the baselines because of the initial exposure provided by DualRec. In other words, ExpoReach can measure the exploration effect. An improvement in the ExpoReach metric means that the creators' new items receive more exposure and feedback, which improves the creator's satisfaction. We measure the improvement of ExpoReach in each single experiment and measure the improvement of DAC in a long-term integrated experiment.

\subsection{Performance of DualRec (Q2)}
\begin{table}
\caption{Performance of DualRec on ExpoReach.}
\vspace{-2mm}
\centering
\begin{tabular}{c|c|c}
    \hline\hline
     \multirow{2}{*}{\textbf{Method}} & \textbf{ExpoReach}& \textbf{ExpoReach} \\
     &\textbf{-1K}&\textbf{-10K} \\
    \hline
    DualRec-v1 v.s. Base & +11.7\%& +8.4\% \\
    \hline
    DualRec-v2 v.s. DualRec-v1 &+4.9\%&+5.4\%\\
    \hline
    DualRec-v3 v.s. DualRec-v2 &+3.2\%&+4.8\%\\
    \hline
    DualRec-v4 v.s. DualRec-v3 &+0.59\%&+0.78\%\\
    \hline
    DualRec-v5 v.s. DualRec-v4 &+3.7\%&+3.5\%\\
    \hline\hline
\end{tabular}
\label{table:performance}
\end{table}

\begin{figure}[t]
    \centering
    \includegraphics[width=0.75\columnwidth, trim=0 10 0 0, clip]{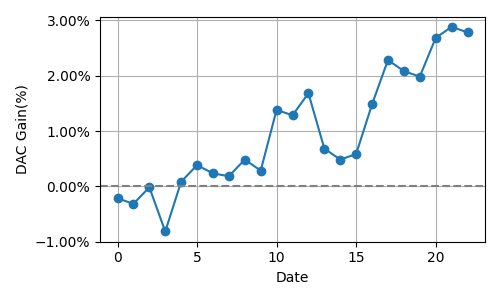}
    \caption{Performance gain of DAC over the baseline.}
    \label{fig:online metrics}
\vspace{-5mm}
\end{figure}

We conduct each experiment for 7 days. Table \ref{table:performance} displays the performance gain of each experiment compared to the previous baseline, with a confidence level greater than 0.95, indicating an improvement in each version. Additionally, we carry out a 21-day experiment to demonstrate the performance gain of the final DualRec algorithm over the baseline, as measured by DAC, as shown in Figure \ref{fig:online metrics}. The DAC gradually improves, reaching a 2.9\% increase by the end of the experimental period. It is shown that every 1\% increase in ExpoReach corresponds to a 0.13\% increase in DAC, indicating that ExpoReach serves as a good proxy metric.

\begin{figure}[t]
    \centering
    \includegraphics[width=0.8\columnwidth,trim=0 6 0 0, clip]{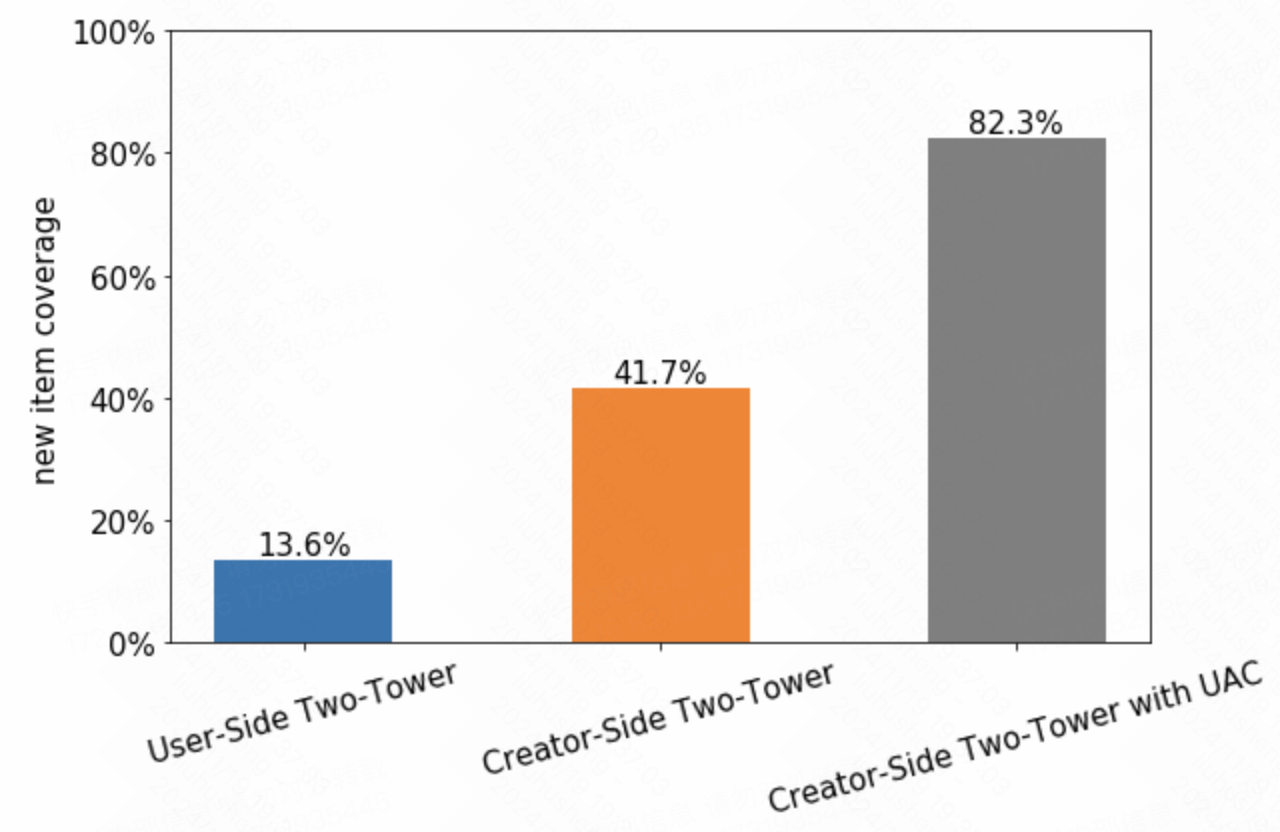}
    \caption{New item coverage of Different Retrievals.}
    \label{fig:coverage-recall}
\vspace{-5mm}
\end{figure}

We also compare the coverage of new items in user-side and creator-side recommendations in Figure \ref{fig:coverage-recall}, where the coverage is defined as the ratio of the number of retrieved items to the total number of new items. This comparison includes i) user-side retrieval with the candidate set restricted to low-exposure new items, ii) the creator-side two-tower model without UAC, and iii) the creator-side two-tower model with UAC. The results show that user-side retrievals exhibit low coverage of new items, even when the candidate set is restricted to these new items. This is due to the tendency of user-side retrievals to select a small proportion of items to enhance user satisfaction. In contrast, creator-side retrievals demonstrate high coverage.

Moreover, all of these experiments enhance the ExpoReach metric. It is important to note that DualRec only matches 200 users for each item, indicating that the increase in future exposure of these items is due to user-side recommendations. This phenomenon is particularly interesting, as DualRec increases the likelihood of these items being discovered by user-side recommender systems.

\subsection{Impacts of UAC Module (Q3)}
The UAC module has been implemented in all versions of DualRec because online improvements cannot be achieved without it. We first discuss the impact of the user activity prediction module, as reflected by the hit rate of matched users returned by the DualRec system. We compare the hit rates with and without the user activity prediction module. When the user activity prediction module is eliminated, DualRec finds users for each item from the entire user set $\mathcal{U}$ rather than from the predicted active user set ${u \in \mathcal{U}: a_u = 1}$. The hit rate is defined as the ratio of matched users who actually come in the following day. The hit rate without the user activity prediction module is only 49\%, while the inclusion of the user activity prediction module raises the hit rate to 82\%, significantly enhancing the effectiveness of the returned user set $\mathcal{U}_i$.

\begin{figure}[t]
    \centering
    \includegraphics[width=0.8\columnwidth,trim=0 0 0 0, clip]{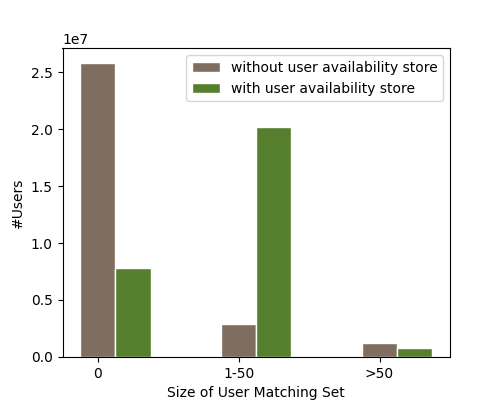}
    \caption{Distribution of user matching set size with and without user availability store.}
    \label{fig:user-availability-store}
\vspace{-3mm}
\end{figure}

Then, we discuss the impact of the user availability store. Figure \ref{fig:user-availability-store} shows the distribution of the user matching set size $\left|\mathcal{D}_u\right|$ with and without the user availability store. Without the user availability store, a small number of users are matched with over 50 items (actually, some receive over 10000 items), exceeding their capacity to watch them, while many users are matched with no items at all. Under this situation, most items cannot be distributed to the users assigned to them, which would prevent DualRec from effectively enhancing creator satisfaction. In contrast, the user availability store ensures that the majority of users are matched with an acceptable number of items, significantly improving matching efficiency. 

\subsection{Impacts of Creator-Side Algorithms (Q4)}
We also test the effectiveness of each part in the DualRec system, as shown in the different versions of DualRec in Table \ref{table:performance}.

\subsubsection{Retrieval Algorithms}
The initial DualRec-v1/v2 includes only the two-tower retrieval, whereas DualRec-v3 and DualRec-v4 introduce the $i2i2u$ and $i2u2u$ retrievals, respectively. The $i2i2u$ retrieval improves the ExpoReach-10K by 4.8\%, while the $i2u2u$ retrieval only increases it by 0.78\%. This difference is due to the $i2i2u$ retrieval capturing video understanding information through the item-similarity calculation module, which is crucial for items with few exposures. In contrast, $i2u2u$ incorporates user similarity,but user similarities are generally less accurate than item similarities, leading to a lower exposure ratio. Table \ref{table:retrieval-performance} presents the exposure ratios for these retrieval methods in the final DualRec system. It shows that the exposure ratios for the two-tower retrieval and the $i2i2u$ retrieval are significantly higher than that of the $i2u2u$ retrieval, consistent with the gains of these retrieval methods.

\begin{table}
\caption{Exposure ratio of different retrievals}
\vspace{-2mm}
\centering
\begin{tabular}{c|c|c|c}
    \hline\hline
    \textbf{Retrievals} & Two-Tower & $i2i2u$ & $i2u2u$\\ 
    \hline
    \textbf{Ratio of Exposure} & 58.7\% & 38.5\% & 2.79\% \\
    \hline\hline
\end{tabular}
\label{table:retrieval-performance}
\vspace{-1mm}
\end{table}
 
\subsubsection{Ranking Algorithms}
We evaluate the impact of the changes discussed in Section \ref{sec:mirror:ranking:prediction} on the prediction model, specifically the sample augmentation (denoted by SA) and the additional features (denoted by EF). Table \ref{table:offline-training} presents the area under the ROC curve (AUC) for different methods, with ``with SA, with EF'' representing the model used in our online experiment. The results indicate that both SA and EF enhance the model's performance. Furthermore, Table \ref{table:performance} shows that DualRec-v2, which incorporates SA and EF in the prediction model, performs significantly better than DualRec-v1, which lacks these modifications. The experimental results also demonstrate the effectiveness of the proposed techniques.

\subsubsection{Score Integration}
DualRec-v5 incorporates modifications to the score integration process discussed in Section \ref{sec:mirror:ranking:score-integration}. Specifically, we define $\theta_u$ as the average interaction rate of user $u$, which appears in the denominator of the integration formula. As a result, the new integration score applies a penalty to users with a high bias. This modification aims to identify users who are genuinely interested in the item, allowing the recommender system to better understand the target users for this item in the future, leading to more accurate recommendations. This adjustment increases the liking rate of items in the experimental groups by 0.78\% and the following rate by 1.472\% while also significantly enhancing the ExpoReach.

\begin{table}
\caption{Offline Results of the Model Changes, where SA denotes sample augmentation, and EF denotes extra features.}
\vspace{-2mm}
\centering
\begin{tabular}{c|c|c}
    \hline\hline
     \textbf{Methods} & \textbf{ AUC} & \textbf{Gain} \\
    \hline
    w/o SA, w/o EF &0.901& 0\\
    w/o SA, with EF &0.908& +0.7pp\\
    with SA, w/o EF &0.911&+1.0pp \\
    with SA, with EF&\textbf{0.913}&\textbf{+1.2pp} \\
    \hline\hline
\end{tabular}
\label{table:offline-training}
\vspace{-2mm}
\end{table}

\section{Conclusion}
This paper introduces a novel creator-side recommender system called DualRec to find the most appropriate users for each item to improve creator satisfaction. We show that the algorithms in traditional user-side recommender systems, including the retrieval algorithms and the ranking algorithms, can be easily translated into the creator-side version with a few changes. Moreover, we discuss the user availability issue in DualRec and introduce UAC to tackle this challenge. We implement DualRec in Kwai, a short video platform with over 100 million users and over 10 million creators, significantly improving creator satisfaction.

\bibliographystyle{ACM-Reference-Format}
\balance
\bibliography{dualrec}

\end{document}